\documentclass[11pt, oneside]{article}   	
\usepackage{geometry}                		
\usepackage{graphicx}				
\usepackage{amssymb}

\usepackage{amsmath}

\graphicspath{{./png/}}

\title{Centennial of General Relativity (1915-2015); The Schwarzschild Solution and Black Holes}
\author{S. M. BLINDER\\
Wolfram Research Inc., Champaign, IL 61820, USA\\ email: sblinder@wolfram.com}
\date{December 2, 2015}							

\begin{document}
\parskip=0.5cm
\parindent=1cm

\maketitle
This year marks the 100th anniversary of general relativity, commemorating the publication of Albert Einstein's paper proposing a set of field equations to describe the structure of spacetime on December 2, 1915 [1]. This theory superseded Isaac Newton's Law of Universal Gravitation, which reigned as a fundamental law of nature for some 250 years. Einstein's theories of relativity have had a subtle but pervasive influence on 20th Century philosophy, art and culture. General relativity ranks as one of mankind's supreme intellectual achievements. Dirac regarded GR as "probably the greatest scientific discovery ever made." Einstein's theory took the form of a complicated set of simultaneous nonlinear partial differential equations, for which exact analytic solutions appeared to be extremely difficult, except for the trivial case of flat spacetime. However, within two months of Einstein's publication, Karl Schwarzschild, a little-known German astrophysicist, produced a remarkably simple exact solution to the field equations for a point mass in otherwise empty space [2]. (Johannes Droste, a student of H. A. Lorentz, independently arrived at the same solution in 1916.)  Einstein was said to have been amazed and delighted. All the more remarkable was the fact that Schwarzschild did his research while serving as an artillery lieutenant in the German army on the Russian front during World War I, in his spare time while computing artillery trajectories. Tragically, while still in military service, he died soon afterward from an autoimmune disease. Further consideration of the Schwarzschild solution later provided the rationale for the existence of black holes, one of the weirdest consequences of GR.  But since the 1960s, which Kip Thorne called the ``golden age of general relativity,'' black holes have become a major field of research in physics and astronomy.

\pagebreak
Figure 1 shows portraits of Einstein and Schwarzschild from the American Institute of Physics collection. Einstein is shown at age 36, when he published the equations of general relativity. 
\begin{figure}
\begin{center} 
\includegraphics[height=10cm]{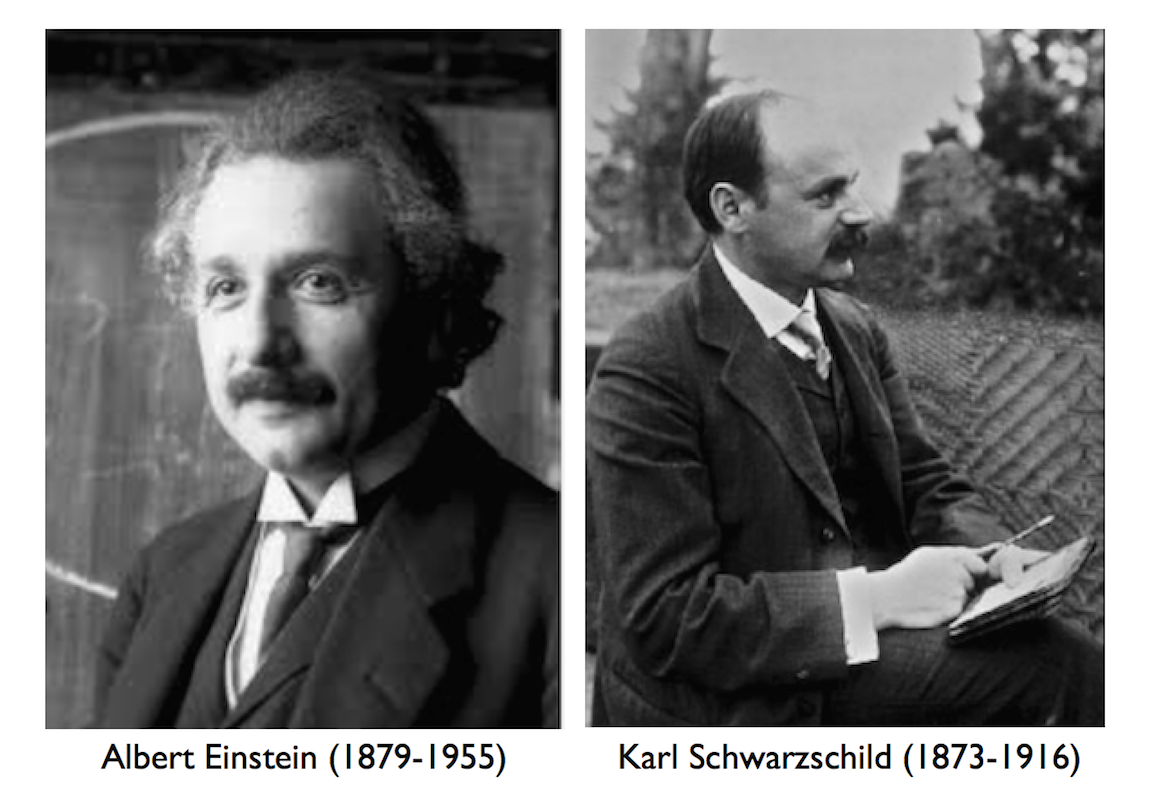}
\end{center} 
\caption{Einstein and Schwarzschild}
\label{einsch}
\end{figure}
Much will be written this year about the seminal contributions of Albert Einstein, so there is no need for us to add any further tributes. We will, instead, focus on Karl Schwarzschild---his original solution and its numerous subsequent elaborations. We begin with the Einstein field equations, which can be written 
$$R_{\mu \nu} - \frac{1}{2} R g_{\mu \nu} = \frac{8\pi G}{c^4} T_{\mu \nu}.$$
Here $G$ is Newton's gravitational constant, $c$, the speed of light, $g_{\mu \nu}$, the spacetime metric, $R_{\mu \nu}$, the Ricci curvature tensor, $R$, the Ricci scalar and $T_{\mu \nu}$, the energy-momentum tensor. (An additional term containing the cosmological constant $\Lambda$ was later added, but does not concern us here.) A qualitative discussion of the components of this equation will suffice. The left-hand side describes the curvature of spacetime, while the right-hand side accounts for the distribution of matter and energy which both produces the curvature and is acted upon by the curvature. As succinctly expressed by John Wheeler, ``Spacetime tells matter how to move; matter tells spacetime how to curve.''  This is to be contrasted to  Newtonian mechanics, in which gravitation is an inverse-square attractive force between masses, described by $F=-G M m/r^2$. Figure 2 contrasts the two different worldviews.
\begin{figure}[h]
\begin{center} 
\includegraphics[height=5cm]{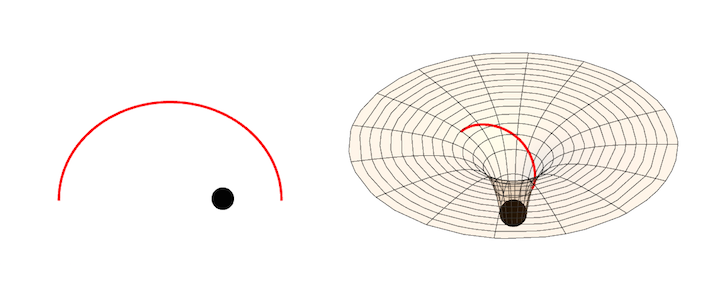}
\end{center} 
\caption{Newtonian gravitation vs curved spacetime}
\label{newton}
\end{figure}
This representation of spacetime curvature is highly schematic. Curvature is actually an intrinsic property of a 4-dimensional continuum. A rough analogy might be the distribution of temperature in an unevenly heated room.

The metric $g_{\mu \nu}$, obtained as a solution of the Einstein field equations, contains a complete description of the geometry of spacetime consistent with some model distribution of matter and energy. This is a generalization of the Pythagorean metric in 3-dimensional Euclidean space: 
$$ds^2 =  dx^2 +  dy^2 + dz^2.$$ The simplest case, free space with no mass or electromagnetic fields, reduces to the Minkowski-Lorentz metric of special relativity,
$$   ds^2 = -c^2 \, dt^2  + dx^2 + dy^2 + dz^2 = -c^2 \, dt^2 + dr^2 +  r^2\, d\theta^2+ r^2 \sin^2 \theta \,  
d\phi^2,$$
where the second form is expressed in spherical polar coordinates. The last two angular terms can be compactly written as $r^2 \, d\Omega^2$. The Schwarzschild solution, representing a point mass $M$ at the origin and free space elsewhere, can be encapsulated in the stationary, spherically-symmetric metric
$$ ds^2 = -\left(1-\frac{2 G M}{c^2 r}\right) c^2 \, dt^2 +\left(1-\frac{2 G M}{c^2 r}\right)^{-1} dr^2 +
r^2 \, d\Omega^2.$$

The standard method to derive this result requires a lengthy computation of the elements of the Ricci tensor along with the stress-energy tensor representing a point mass $M$ at the origin. This confirms that the Schwarzschild solution satisfies the Einstein field equations. The details are available in many references. We can, more simply, rationalize the Schwarzschild solution by a heuristic argument which avoids all the tedious tensor algebra of general relativity. Consider a test mass $m << M$ in free fall with an instantaneous radial velocity $v$, in a comoving Lorentz frame described by the instantaneous variables $t'$ and $r'$, with the effective metric  $ds^2 =-c^2 \, dt'^2+dr'^2$. To a stationary observer, the falling body exhibits a time dilation, with  $dt=dt'/\sqrt{1-v^2/c^2}$, and a Lorentz-FitzGerald contraction, such that $dr=dr' \, \sqrt{1-v^2/c^2}$. Thus the metric for the stationary observer can be expressed as
$$          ds^2 =-\left(1-\frac{v^2}{c^2}\right) c^2 \,dt^2 +\left(1-\frac{v^2}{c^2}\right) ^{-1} dr^2+ r^2 \, 
d\Omega^2.  $$
Finally, we identify $v$ with the free-fall velocity (equal in magnitude to the escape velocity) in Newtonian mechanics. This occurs when the total energy equals zero, with the kinetic energy of the mass $m$ balancing its potential energy, such that  $\frac{1}{2} m v^2-(G M m)/r=0$. Therefore $v=\sqrt{2G M /r}$. And substituting this value into the above metric gives the Schwarzschild result. The Schwarzschild radius can be defined by  $r_S = 2G M /c^2$. The sphere of radius $r_S$ is called the event horizon. Setting $c=1$, the metric can be compactly expressed  
$$    ds^2 =-\left(1-\frac{r_S}{r}\right) dt^2+\left(1-\frac{r_S}{r}\right)^{-1} dr^2+ r^2\, d \Omega^2 $$

When $r > r_S$, any motion is timelike, with $ds^2 < 0$, and conforms to the usual ordering of cause and effect. However, inside the Schwarzschild radius, where $r < r_S$, the metric becomes spacelike, with 
$ds^2 > 0$, and $r$ effectively becomes a time coordinate. As will be shown below, no outgoing light or material object can cross the event horizon. In physical terms, for $r<r_S$, spacetime approaches infinite curvature, such that the force of gravity becomes so irresistible that the escape velocity becomes greater than the speed of light; but nothing can go faster than light. Thus, when an object is inside the horizon, it is trapped forever. We have a black hole! The term is usually credited to John Wheeler in 1967. As early as the 18th Century, it had been speculated (John Mitchell, Pierre-Simon Laplace) that a body might become so massive that even light could not escape from it. It is now understood that after a star exhausts its nuclear fuel, it can collapse into a white dwarf, or even further into a neutron star. Around 1930, Subrahmanyan Chandrasekhar derived the result that a star more than 1.4 times as massive as the Sun can collapse even further into a black hole. We limit our consideration to nonrotating, uncharged black holes, with a point mass localized at $r=0$. Moreover we are limiting ourselves to classical general relativity, with no account is taken of quantum effects. A Schwarzschild black hole is assumed to be a permanent structure. However, more recently, Stephen Hawking took account of the effect of extremely strong fields according to quantum field theory. He proposed that virtual particle-antiparticle pairs are spontaneously created near the event horizon. Should one particle of a pair be swallowed by the black hole, it would appear that its companion particle is being emitted by the black hole. Thus black holes do, in effect, emit radiation and they can eventually ``evaporate.''

The locus of $r=0$ is designated as the singularity of a black hole. At a singularity the known laws of physics are likely to break down. There is perhaps the possibility of some type of phase transition. The singularity will turn out to have a structure much more complex than just a point. The event horizon, a sphere of radius $r_S$  around the singularity, is a boundary in spacetime beyond which events inside cannot affect an observer outside. It might be thought of as some sort of ``shield'' protecting the outside world from what goes on inside the black hole. Interestingly, the translation of the German word {\it Schwarzschild} is ``black shield.''

A casual view of the Schwarzschild metric might imply that the event horizon, with radius $r=r_S$, is also a singularity, since $(1-r_S/r)^{-1} \to \infty$ as $r \to r_S$. However, it was recognized rather early (Eddington, Lema\^itre) that this is just an artifact of the coordinate system---for example, in spherical polar coordinates, the irregular behavior of the spherical angle $\phi$ on the polar axis. The event horizon separates the pseudo-Riemannian manifold representing spacetime into interior and exterior sub-manifolds. There is no limitation on signals sent from the exterior to the interior, but the reverse direction is not possible. In other words, the interior of the black hole is causally disconnected from the rest of the universe. It is suggestive that analytic continuation of the Schwarzschild solution might cover the interior manifold, as well as revealing the existence of additional regions of spacetime.

A definitive extension of the Schwarzschild structure of spacetime was worked out in 1960, independently by Martin Kruskal and by George Szekeres [3]. We will describe a simplified, watered-down version of their model, mainly emphasizing its qualitative aspects. A useful starting point is to find the null geodesics, the paths of light rays, of the Schwarzschild metric, along which $ds=0$. Limiting ourselves to radial trajectories (with 
$d\Omega=0$), we find
$$       -\left(1-\frac{r_S}{r}\right) dt^2+\left(1-\frac{r_S}{r}\right) ^{-1} dr^2=0 \qquad
   \mbox{which gives}\qquad   \frac{dr}{dt}=\pm\left(1-\frac{r_S}{r}\right). $$
In terms of Regge-Wheeler tortoise coordinates, defined by  $r^*(r)=r + r_S\, \log|r/r_S-1|$, incoming and outgoing null geodesics for $r>r_S$ can be written $t+r^*(r)=v$ and $t-r^*(r)=u$, respectively, where $v$ and $u$ are constants. For an incoming light ray, when $r>>r_S$, $r$ decreases nearly linearly with $t$. But as
 $r \to r_S$,  $t \to \infty$, meaning that the light ray begins to crawl increasingly slowly as it approached the horizon, so that it would take an infinite time to actually reach the horizon. (This is evidently reminiscent of Zeno's paradox about Achilles and the Tortoise.) Actually, in the proper time of an infalling object, it will pass through the horizon without any unusual effects and fall directly into the singularity. Only a stationary observer will see the object lingering forever around the horizon. It has been proposed therefore that all the particles that enter a black hole might leave traces of their information imprinted somehow on the event horizon. This resembles a hologram in the sense that 3D information is recorded on a 2D surface. Gerard 't Hooft and Leonard Susskind have even speculated that the information content of the entire Universe can be represented by a hologram on the cosmological horizon.

In the interior of the event horizon, where $r<r_S$, the formulas for incoming and outgoing null geodesics are reversed. The incoming object starts at $r=r_S$ at the time $t=-\infty$ and is swallowed by the singularity at $r=0$, when $t=0$. Again, for an observer falling with the object, the trajectory is simply a continuation of its  $r>r_S$ motion and is completed in a finite interval of proper time. For the outgoing geodesic, we find again that $r=0$, when $t=0$. But as $t$ increases toward $+\infty$, $r$ approaches a boundary of radius $r_S$, and connects with outgoing geodesics for $r>r_S$. This can not be the same event horizon as above, since nothing inside it can escape. The penetrable boundary is called an {\it antihorizon}. Thus we have a singularity that spontaneously expels all of its contents to the outside world. This is called a {\it white hole}, which implies an outlet into a new region of spacetime, possibly a whole new world. This might belong to a distant part of our Universe, or perhaps to another parallel Universe. The passage between the two worlds is variously called an {\it Einstein-Rosen bridge} or a {\it wormhole} (a term also coined by John Wheeler). It should be emphasized that white holes and wormholes, although suggested by solutions of the equations of general relativity, have never been found in the real world and might just be mathematical artifices (although such tunnels are much beloved in science fiction). By contrast, there is ample evidence for the existence of black holes.

\begin{figure}
\begin{center} 
\includegraphics[height=7cm]{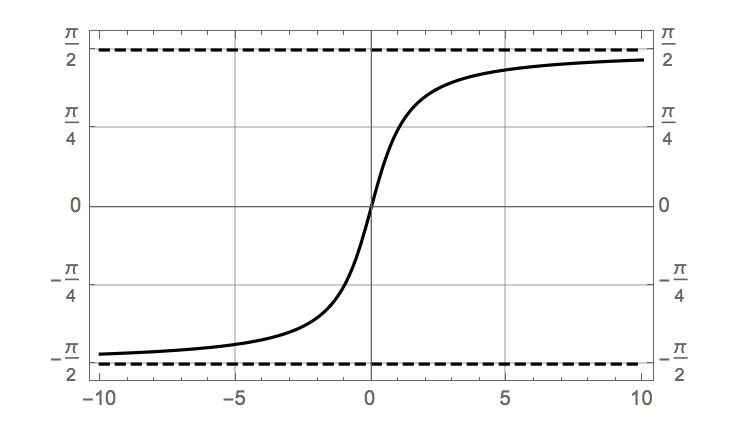}
\end{center} 
\caption{Plot of $\arctan(x)$}
\label{arctan}
\end{figure}

\begin{figure}
\begin{center} 
\includegraphics[height=7cm]{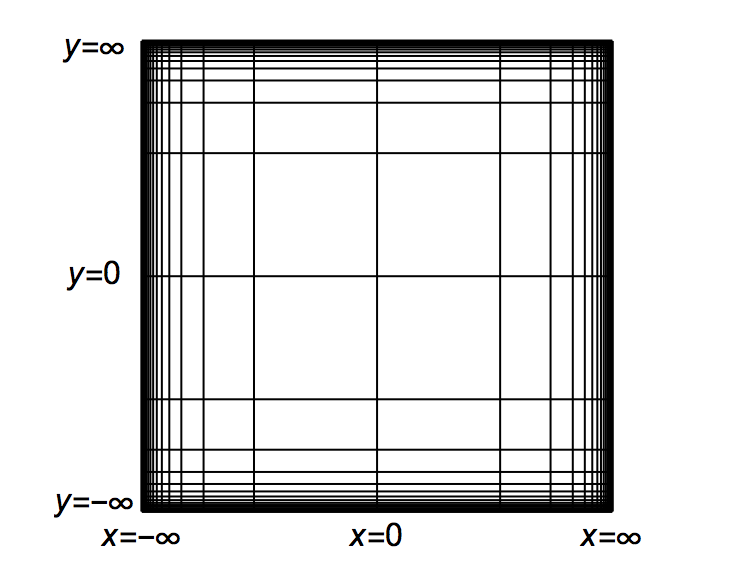}
\end{center} 
\caption{Compact representation of the 2D-Cartesian plane}
\label{plane}
\end{figure}

The structure of 4-dimensional spacetime can be represented by diagrams, which are, in a sense, two-dimensional cross sections. The vertical axis is usually time and the horizontal axis is one of the space coordinates. The latter can be a radial coordinate, so that each value of $r$ can be pictured as representing a 3-dimensional sphere. Penrose (or Carter-Penrose) diagrams provide very lucid pictorial representations of the complete causal structure of any spacetime geometry. They are indispensable for navigating around black holes. Penrose diagrams are compact in space and time variables, so that points at infinity (in space or time) appear as finite points or lines in the diagram. This is made possible by the unique behavior of the arctangent function: for $x$ varying from $-\infty$ to $+\infty$, $\arctan(x)$ has the finite limits $\pm\pi/2$, as shown in 
Figure \ref{arctan}. 
The two-dimensional Cartesian plane, scaled using $\arctan(x)$ and $\arctan(y)$ allows the entire infinite plane $-\infty < x, y < \infty$ to be plotted in a finite region, as shown in Figure \ref{plane}.

\newpage
On Penrose diagrams, null geodesics, the paths of light rays, are lines at $\pm 45^\circ$ angles, in common with other spacetime pictures. Figures 5, 6 and 7 show Penrose diagrams for Minkowski space, Schwarzschild black holes and Kruskal-Szekeres maximally extended spacetime, respectively.
For Minkowski spacetime, the radial variable $r$ and time $t$ are related to the null geodesics $u$ and $v$ by 
$r \pm t=\tan(u \pm v)$. The Penrose diagram is a diamond-shaped figure with each corner representing an infinite limit point of space or time. The two upper sides can be associated with future lightlike infinity, the two lower sides, to past lightlike infinity. The diagram for a Schwarzschild black hole builds on the Minkowski diagram, adding a wedge-shaped region with boundaries at the event horizon and the antihorizon, both spheres of radius $r=r_S$, and the black hole singularity, represented by a jagged line with $r=0$.  Extended spacetime, shown in Figure 7, is seen to consist of four regions, including a duplicate copy of the Schwarzschild geometry, but reversed in time and connected along the antihorizon. The analytic extension contains not only our universe (Region I) and a black hole (Region II), but also a parallel universe (Region III), connected by an Einstein-Rosen bridge, and a white hole (Region IV). This is actually all the product of mathematical construction; any of its predictions may or may not have a basis in reality.
\begin{figure}
\begin{center} 
\includegraphics[height=7cm]{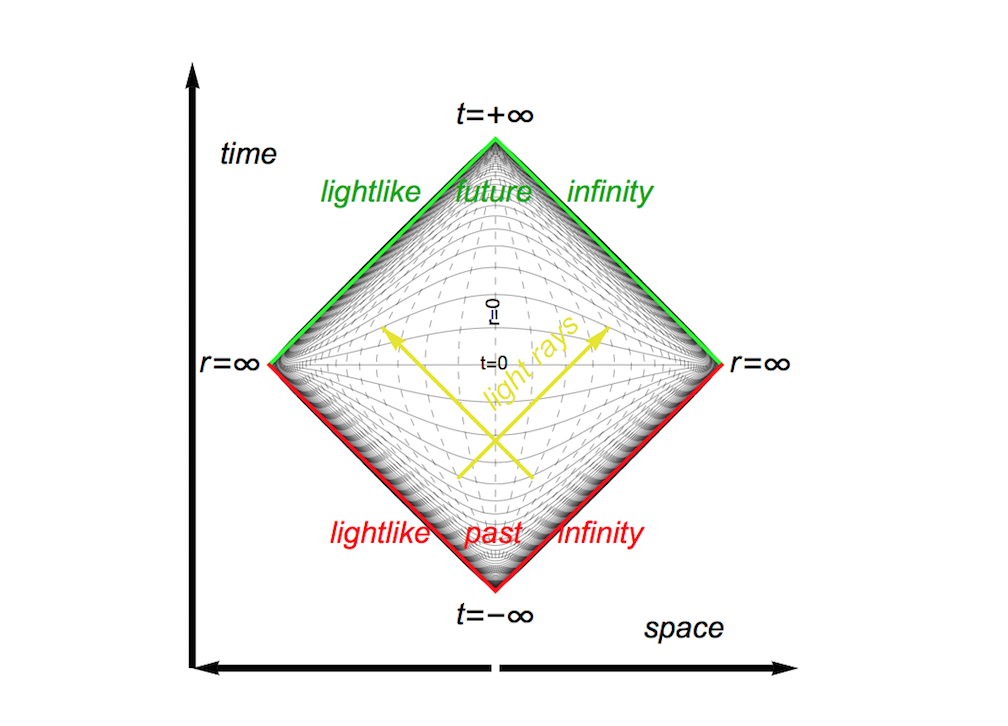}
\end{center} 
\caption{Penrose diagram of Minkowski space}
\label{mink}
\end{figure}
\begin{figure}
\begin{center} 
\includegraphics[height=7cm]{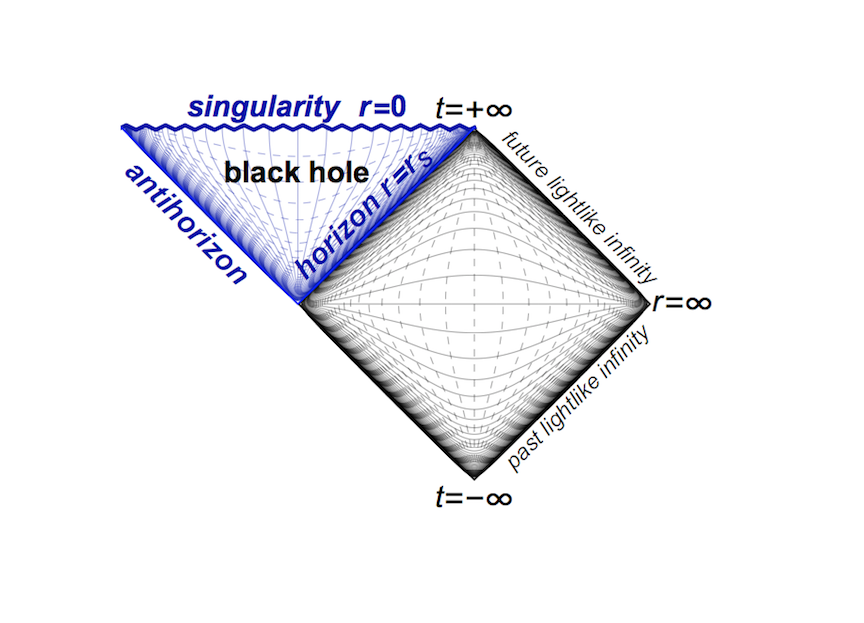}
\end{center} 
\caption{Penrose diagram for Schwarzschild black hole}
\label{sch}
\end{figure}
\begin{figure}
\begin{center} 
\includegraphics[height=7cm]{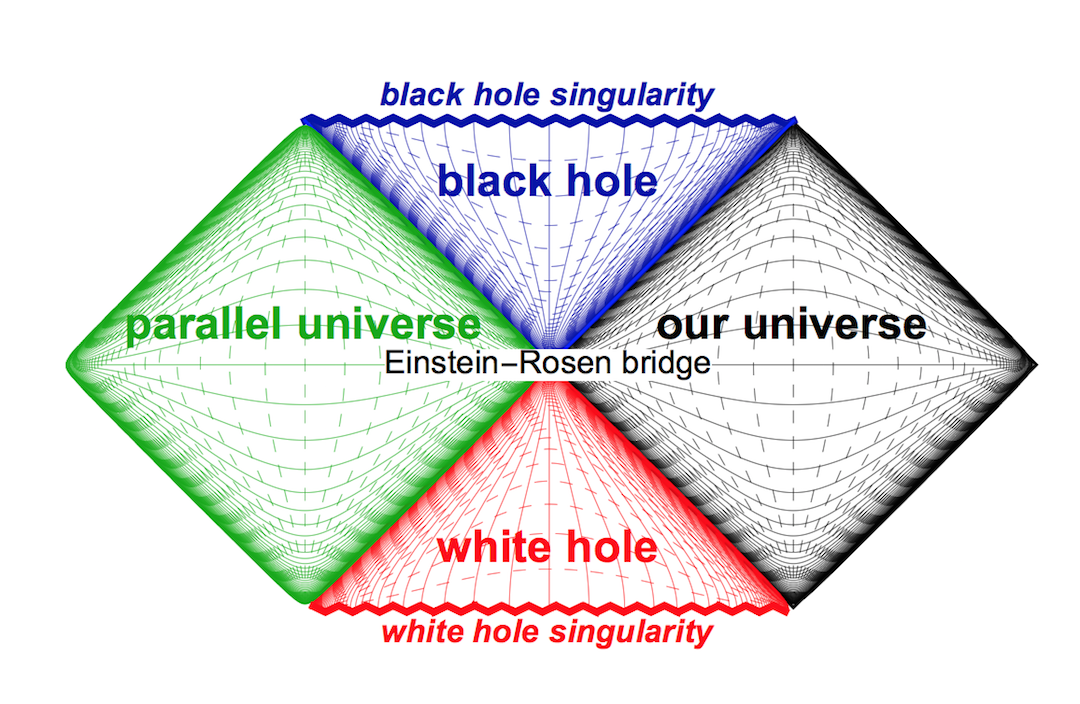}
\end{center} 
\caption{Penrose diagram for Kruskal-Szekeres extended spacetime}
\label{ext}
\end{figure}

\newpage
Thus, as we have seen, during the past century, the first nontrivial exact solution of Einstein's field equations, derived by an obscure academic, Karl Schwarzschild, has evolved into the vast and complex study of black holes. With the insight of Stephen Hawking, black holes have become an intersection of general relativity and quantum theory (as well as thermodynamics) and might provide significant clues to their eventual unification into a ``Theory of Everything.''

\vspace{2 cm}

\noindent References:

\noindent [1] A. Einstein: Die Feldgleichungen der Gravitation, Sitzungsberichte der Preussischen Akademie der Wissenschaften zu Berlin, 1915 (Part 2), 844-47.

\noindent [2] K. Schwarzschild: \"Uber das Gravitationsfeld eines Massenpunktes nach der Einsteinschen Theorie, Sitzungsberichte der K\"oniglich Preussischen Akademie der Wissenschaften,1916, 7, 189-196. 

\noindent [3] C. W. Misner, K. S. Thorne, J. A. Wheeler, 1973, Gravitation, W. H. Freeman and Co.

\noindent See also the September 2015 Special Issue of Scientific American, which commemorates 100 years of general relativity.

\end{document}